# Infinite-volume mixing for dynamical systems preserving an infinite measure

## Marco Lenci

*Dipartimento di Matematica, Università di Bologna, Piazza di Porta San Donato 5, 40126 Bologna, Italy*
*Email: marco.lenci@unibo.it*

### Abstract

In the scope of the statistical description of dynamical systems, one of the defining features of chaos is the tendency of a system to lose memory of its initial conditions (more precisely, of the distribution of its initial conditions). For a dynamical system preserving a probability measure, this property is named 'mixing' and is equivalent to the decay of correlations for observables in phase space. For the class of dynamical systems preserving infinite measures, this probabilistic connection is lost and no completely satisfactory definition has yet been found which expresses the idea of losing track of the initial state of a system due to its chaotic dynamics. This is actually on open problem in the field of infinite ergodic theory. Virtually all the definitions that have been attempted so far use *local observables*, that is, functions that essentially only "see" finite portions of the phase space. In this note we introduce the concept of *global observable*, a function that gauges a certain quantity throughout the phase space. This concept is based on the notion of *infinite-volume average*, which plays the role of the expected value of a global observable. Endowed with these notions, whose rigorous definition is to be specified on a case-by-case basis, we give a number of definitions of *infinite mixing*. These fall in two categories: global-global mixing, which expresses the "decorrelation" of two global observables, and global-local mixing, where a global and a local observable are considered instead. These definitions are tested on two types of infinite-measure-preserving dynamical systems, the random walks and the Farey map.

## 1. Introduction

In ergodic theory—the mathematical discipline that studies the stochastic properties of dynamical systems—one of the main features (some will say the defining feature) of a chaotic system is the *mixing property*. Specializing to discrete time for simplicity, we say that the dynamical system given by the map $T{:}M{\to}M$, preserving the normalized measure $\mu$ on $M$, is mixing if, for all (measurable) subsets $A,B \subseteq M$,

$$\lim_{n \to \infty} \mu(T^{-n}A \cap B) = \mu(A)\mu(B). \tag{1}$$

In other words, the portion of $T^{-n}A$ within $B$, in the sense of the measure, tends, as $n{\to}\infty$, to the portion of $T^{-n}A$ within the whole $M$ (because $\mu(T^{-n}A){=}\mu(A)$ and $\mu(M){=}1$); whence the term 'mixing'.

The condition that the system be endowed with a dynamics-invariant measure is very important, but not particularly restrictive. Many systems of physical interest (e.g., all Hamiltonian systems) come with a predefined, physically relevant, invariant measure. For all others, theorems exist that guarantee the existence of an invariant measure [1].

What is pivotal here is the assumption that $\mu$ be finite, in which case one assumes, with no loss of generality, that $\mu(M){=}1$. It is easy to see that (1) is equivalent to the following: For all square-integrable functions (a.k.a. *observables*) $f,g{:}M{\to}\mathbf{R}$,

$$\lim_{n \to \infty} \int_M (f \circ T^n)g \, d\mu = \int_M f \, d\mu \int_M g \, d\mu. \tag{2}$$

Namely, if we think of observables as *random variables* defined on the probability space *(M,μ)*, (2) reads: the evolution of the observable $f$ (that is, $f{\circ}T^n$) *decorrelates* with the observable $g$, as time goes to infinity. And so mixing means 'decay of correlations'. In other words, a mixing dynamical system is one for which the iteration of the dynamics makes the system lose memory of its initial conditions.

This shows that mixing is an intrinsically probabilistic concept. But what if $\mu$ is an infinite measure? The probabilistic interpretation no longer makes sense, but one can certainly think of chaotic systems defined on a phase space of infinite measure; for example, an open dynamical system such as an unbounded Lorentz gas, or a system on $N$ particles in celestial mechanics, etc. One expects these systems and many others to disperse, or "mix up", trajectories in a way that is analogous to our description of (1). Thus, it is natural to look for a good definition of mixing for dynamical systems preserving an infinite measure (for short, *infinite mixing*).

In fact, this has been an open problem for over 70 years, at least judging from the first reference to such questions in the scientific literature, which seems to be due to Hopf in 1937 [2]. We now give a scanty outline of the history of the problem, referring the interested reader to [3] and especially the bibliography therein.

Hopf touched upon the question of infinite mixing by means of an example in his celebrated textbook of ergodic theory [2]. It consists of a half-strip $M$ in the plane and a map $T{:}M{\to}M$ that preserves the Lebesgue measure on *M,* denoted $\mu$. Hopf shows that there exists an increasing sequence $\{\rho_n\}$ of positive numbers such that, for all finite-measure sets $A,B \subseteq M$ with zero-measure boundaries,

$$\lim_{n \to \infty} \rho_n \, \mu(T^{-n}A \cap B) = \mu(A)\mu(B). \tag{3}$$

Without further elaboration, he calls the triple *(M,μ,T)* 'mixing'. Comparing (3) with (1), we notice that the left-hand side is multiplied by the ever-larger number $\rho_n$. In fact, one expects that, for all reasonable sets $A$ and $B$, the portion of $T^{-n}A$ within $B$ vanishes, as $n{\to}\infty$, because $T^{-n}A$ has infinite room to spread through. Therefore, one can only observe the mixing if the left-hand side of (1) is rescaled by an appropriate amount, which depends on the system but not on the pair of sets under scrutiny. For this reason, in current parlance, $\{\rho_n\}$ is referred to as *the scaling rate* [4]. Finally, notice that (3) cannot hold for *all* finite-measure sets $A,B$, as derived by a result of Hajian and Kakutani [5].

In 1967 Krickeberg proposed (3) as *the* definition of infinite mixing, at least for the general case where $T$ is an almost-everywhere continuous map of a Borel space $(M,\mu)$ into itself, preserving $\mu$ [6]. (There are other, more technical but less essential, conditions that we do not discuss here.) Krickeberg's definition has been studied and applied several times in the literature (sometimes without even knowing it), but not to the point of being universally accepted—or known.

There is another couple of definitions of infinite mixing that were given at the end of the 1960's and have had a certain degree of popularity since then. They are due to Krengel and Sucheston [7] and are based on the discovery that *finite* mixing in ordinary ergodic theory can be restated as follows: for every measurable set $A \subseteq M$, the sequence $\{T^{-n}A\}$ is *semiremotely trivial* with respect to $\mu$ [8]. (What semiremotely trivial means is not essential here and we refer the interested reader to [8] or [7].) Motivated by the intention of giving definitions of a purely measure-theoretical nature, Krengel and Sucheston called a $\mu$-non-singular map $T:M \rightarrow M$ *mixing* if $\{T^{-n}A\}$ is semiremotely trivial for all $A$ with $\mu(A)<\infty$, and *completely mixing* if the same happens for all measurable $A \subseteq M$ ($\mu$-non-singular means that $T$ cannot map a positive-measure set, relative to $\mu$, into a null-measure set).

Specializing to the measure-preserving case, it is proved in [7] that a large class of perfectly reasonable maps (including all invertible ones) cannot be completely mixing and that mixing is equivalent to

$$\lim_{n \to \infty} \mu(T^{-n}A \cap B) = 0,$$

for all finite-measure sets $A,B$. This is a much too weak version of (3) and cannot be considered satisfactory, because it classifies too many systems—for example, the translations of a Euclidean space—as mixing.

Although, as mentioned, these definitions have been discussed and applied by the scientific community, the problem of finding a definition of infinite mixing as general and uncontroversial as (1)-(2) for finite mixing remains open—and indeed unsolvable for some [9].

Apart from Krengel and Sucheston's definition of complete mixing, which we have disqualified immediately, what all these notions have in common, among themselves and with some very refined recent work on infinite mixing [10-11], is that they involve only finite-measure sets, or, if we prefer to use observables rather than sets, only integrable or square-integrable functions. In the opinion of this author, this is the most serious shortcoming in all the past research on the question of infinite mixing.

In this note we hope to convince the reader that a new concept is needed, a way to observe the *global* state of the whole dynamical system, at once. In the terminology of [3], we need to introduce *global observables*, and a way to measure them, analogous to the integral (i.e., the expectation of) *local observables* in finite ergodic theory.

To do that, we will resort to a notion that has been extremely useful in statistical mechanics since its inception. Statistical mechanics deals with systems that are very large, compared to the microscopic scale of its constituents, and describes them mathematically as infinite systems. In order to describe one aspect, one experimental measure, for any such system, one associates to the measured quantity one "extensive observable" (that is, one function of the phase space that behaves essentially in the same way in every region of the space; e.g., the energy, if the measured quantity is the temperature), computes its mathematical expectation (that is, an averaged integral) on a large portion $V$ of the phase space, and finally takes the limit as $V$ covers the whole space. This process is called, depending on the details of the limit, *thermodynamic limit* or *infinite-* *volume limit*. Here we choose the second expression, which is less charged with experimental significance.

In Section 2 we will introduce the concept of global observable as a function $F:M \to \mathbf{R}$, together with its infinite-volume average $A_\mu(F)$. This is a notion that requires a choice, depending on the system at hand, of what is to be considered a large set in phase space, and what quantities one wants to measure there. Once these choices have been made, rigorous definitions of infinite mixing can be given. They will be variations and specifications of the following two limits.

$$\lim_{n \to \infty} A_\mu((F \circ T^n)G) = A_\mu(F)\, A_\mu(G); \tag{4}$$

$$\lim_{n \to \infty} \mu((F \circ T^n)g) = A_\mu(F)\, \mu(g), \tag{5}$$

where $F, G$ are two global observables and $g$ a *local observable*, say an integrable function, on $M$. We have used the customary notation $\mu(g) = \int g\, d\mu$.

We call the first type of definition *global-global mixing*, as it somehow expresses the decorrelation between two global observables; and the second type *global-local mixing* because the coupling is between a global and a local observable.

The interpretation of (5) can be particularly revealing. Suppose one takes $g \geq 0$ and $\mu(g)=1$. Then $\mu_g$, the probability measure defined by $d\mu_g = g\, d\mu$, can be considered an initial state for the system, in a stochastic sense. The left-hand side of (5) then reads $\mu_g(F \circ T^n)$, or $Q^n \mu_g(F)$, where $Q$ is the evolution operation, relative to $T$, for the probability measures on $M$ (in technical terms, it is the version of the Perron-Frobenius operator of $T$ that acts on the $\mu$-absolutely continuous finite measures on $M$). So (5) says that $Q^n \mu_g$ "converges weakly" to $A_\mu$, when applied to global observables. Therefore, $A_\mu$ acts as a sort of equilibrium state for the system.

In Section 3 we give two examples of infinite-measure-preserving dynamical systems on which the new definitions can be tested, the random walk and the Farey map. In Section 4 we present our mathematical results, both general sufficient conditions for some of our definitions of infinite mixing, and specific results for the systems of Section 3.

## 2. Global and local observables and the definitions of infinite mixing

In the rest of this note, we call 'dynamical system' the triple *(M, μ, T)*, where $M$ is a measure space endowed with the measure $\mu$, and $T$ is a $\mu$-preserving (measurable) map $M \to M$, i.e., $\mu(T^{-1}A) = \mu(A)$, for all (measurable) $A \subseteq M$. Sometimes it necessary to specify what the measurable sets are, that is, to specify the $\sigma$-algebra $\Sigma$ defined on $M$. In that case we present the dynamical system as the quadruple *(M, Σ, μ, T)*.

We assume that $\mu(M) = \infty$ and, to avoid pathological situations, that *μ is σ-finite* (which means that there is an increasing sequence of sets $M_n \subset M$, with $\cup_n M_n = M$, such that $\mu(M_n) < \infty$ [9]). This setup accommodates both invertible and non-invertible maps. For dynamical systems with continuous time (flows), generalizing the forthcoming ideas, definitions and results is straightforward, cf. [3].

### 2.1. The exhaustive family

In order to put the notion of "large set" on formal grounds, we posit the existence of an *exhaustive family*. An exhaustive family $\mathcal{V}$ is a collection of finite-measure subsets of $M$ that contains at least one increasing sequence $\{V_n\}$ (w.r.t. the inclusion) such that $\cup_n V_n = M$. By definition, there will always be infinitely many ways to choose an exhaustive family. In some cases, different choices may lead to dramatically different result. The right choice is a decision for the researcher to make, depending on what he/she considers a large set to be, in the system at hand. More on this point later. In any case, $\mathcal{V}$ must verify at least the following assumption.

**(A1)**     As $V \uparrow M$, $\mu(T^{-1}V \Delta V) = o(\mu(V))$, uniformly in $V$ belonging to $\mathcal{V}$.

(Let us recall that the *symmetric difference* of two sets $A$ and $B$ is defined as $A \Delta B = (A \cup B) \backslash (A \cap B)$. Clearly if $\mu(A \Delta B) = 0$, the sets $A$ and $B$ differ only by a null set.)

The physical meaning of the above is rather intuitive. If $V$ is a very large set form our exhaustive family, whose role is to represent somehow the whole of $M$, a finite-time application of the dynamics should not change it by much. In other words, at least in most parts of the phase space, the dynamics should act on a scale that is much smaller that any large set of $\mathcal{V}$. Other than being physically sound, **(A1)** also has an important mathematical consequence, cf. Proposition 1 below.

Given $\mathcal{V}$ and a function $F : M \rightarrow \mathbf{R}$, we define the *$\mu$-uniform infinite-volume average of $F$ w.r.t. the exhaustive family $\mathcal{V}$*—in short, the *infinite-volume average of $F$*—as

$$A_\mu(F) = \lim_{V \uparrow M} \frac{1}{\mu(V)} \int_V F \, d\mu, \tag{6}$$

when the limit exists. The limit $V \uparrow M$ means that, for every $\varepsilon > 0$, a number $N = N(\varepsilon)$ can be found such that, for all $V$ in $\Sigma$ with $\mu(V) > N$, the right-hand side of (6) must be $\varepsilon$-close to the left-hand side. (So there is uniformity in the position of $V$, a very important requirement, see [3] and Section 3.3.

## 2.2. Global observables

Once the choice has been made of *where* to make our measurements, that is, once we have chosen the collection of large sets $\mathcal{V}$, the choice must be made of *what* to measure, that is, what observables we want to think as random variables defined on ever-larger sets $V$ from $\mathcal{V}$. In other words, one needs to define the class $\mathcal{G}$ of the global observables. Just like $\mathcal{V}$, this is a human-made choice and reflects the type of measurements the researcher wants to—or can—describe mathematically. Generally speaking, these functions must have a non-negligible support throughout the phase space and be qualitatively similar in different regions of it. Put it suggestively, they should "measure something at infinity".

The choice of $\mathcal{G}$ is even more delicate than that of $\mathcal{V}$. If $\mathcal{G}$ is too small, the forthcoming definitions of mixing might not be very meaningful (one might have excluded from $\mathcal{G}$ a completely reasonable subclass of observables that falsify (4) or (5)). If $\mathcal{G}$ is too large, the definitions might be too hard to verify, or even false, even though everyone would agree the system in question should be regarded as mixing. We believe there is a happy medium in most cases, certainly in the examples that we give in Section 4.

Whatever the choice of $\mathcal{G}$, a minimal set of assumptions must be verified, in order for the whole scheme to make mathematical sense.

**(A2)**     All functions of $\mathcal{G}$ are essentially bounded (i.e., they are bounded except on $\mu$-null sets).

**(A3)**     $A_\mu(F)$, as defined by (6), exists for all functions $F$ in $\mathcal{G}$.

The need for **(A3)** is self-evident. As for **(A2)**, we will see later that the boundedness of the global observables is the most natural condition to require, if certain integrals are to exist, cf. **(A4)** later on.

A consequence of **(A1)-(A3)** is the invariance of the infinite-volume average for the dynamics, which will be fundamental in our definitions of infinite mixing.

**Proposition 1.** If $F$ belongs to $\mathcal{G}$, $A_\mu(FoT^n)$ exists and equals $A_\mu(F)$, for all $n$.

*Proof.* It suffices to show the above for $n=1$ and apply the result multiple times. Using **(A1)-(A2)** and the invariance of $\mu$, we get

$$\lim_{V\uparrow M} \frac{1}{\mu(V)} \int_V F \circ T \, d\mu = \lim_{V\uparrow M} \frac{1}{\mu(V)} \int_{T^{-1}V} F \circ T \, d\mu + o(1) = \lim_{V\uparrow M} \frac{1}{\mu(V)} \int_V F \, d\mu + o(1),$$

whence the result follows via **(A3)**. Q.E.D.

### 2.3. Local observables

We also need observables that are in a sense opposite—better yet, dual—to the global observables. They represent microscopic measurements in a macroscopic system and are given by functions that, morally, are supported on finite-measure sets. We call them *local observables* and denote their class by $\mathcal{L}$. As for $\mathcal{V}$ and $\mathcal{G}$, choosing $\mathcal{L}$ is a prerogative of the person who studies a given dynamical system. In this case, however, the choice is much less delicate than that of $\mathcal{G}$. This should become clear below.

$\mathcal{L}$ has a minimal requirement too:

**(A4)**   All functions of $\mathcal{L}$ are integrable with respect to $\mu$.

Here and in the rest of the paper we indicate global observables with upper-case Roman letters and local observables with lower-case Roman letters.

Given the dynamical system $(M,\mu,T)$, a suitable exhaustive family $\mathcal{V}$, and the classes $\mathcal{G}$ and $\mathcal{L}$ of global and local observables, the two notions of mixing described in (4)-(5) can be put on solid mathematical grounds. We give two definitions of global-global mixing and three definitions of global-local mixing, and explain the significance of the several versions.

### 2.4. Global-global mixing

The two definitions of global-global mixing are as follows:

**(M1)**   If $F,G$ belong to $\mathcal{G}$, then

$$\lim_{n\to\infty} A_\mu((F \circ T^n)G) = A_\mu(F) A_\mu(G).$$

**(M2)**   If $F,G$ belong to $\mathcal{G}$, then

$$\lim_{\substack{n\to\infty \\ \mu(V)\to\infty}} \frac{1}{\mu(V)} \int_V (F \circ T^n) G \, d\mu = A_\mu(F) A_\mu(G).$$

(The above limit means that, for every $\varepsilon>0$, a number $N=N(\varepsilon)$ can be found such that, for all $V$ in $\Sigma$ with $\mu(V)>N$ and all $t>N$, the above left-hand side is $\varepsilon$-close to the right-hand side.)

**(M1)** is perhaps the most natural definition one can come up with if they want to describe the "decorrelation" of two global observables (notice that Proposition 1 is crucial for this interpretation). However, its left-hand side might be ill-posed, even before the limit in $n$ is taken: in fact, the existence of the infinite-volume average of $F$ (hence $FoT^n$) and $G$ in no way implies the same for *(FoTⁿ)G*. This would amount to a ring property for $\mathcal{G}$ that we have not postulated—nor should we, lest $\mathcal{G}$ become too small and rigid.

One way to work around this problem—other than declaring that when $A_\mu((FoTⁿ)G)$ does not exist, the limit in **(M1)** is automatically false—is to devise another definition, along the same line as **(M1)** but with no question of well-posedness: this is exactly what **(M2)** achieves. A dynamical system is **(M2)**-mixing if, for every pair of global observables, the evolution of one observable is almost uncorrelated with the other observable, on all large sets $V$ and for all large times $n$. The limit in $V$ is uniform with respect to $n$ and viceversa. This last point causes **(M2)** to be in some sense stronger than **(M1)**, as we will see in more detail in Proposition 2 below.

## 2.5. Global-local mixing

As for global-local mixing, here are the three definitions:

**(M3)**    If $F$ belongs to $\mathcal{G}$ and $g$ belongs to $\mathcal{L}$, with $\mu(g)=0$ (i.e., $g$ is a zero-average local observable), then

$$\lim_{n \to \infty} \mu((F \circ T^n)\,g) = 0.$$

**(M4)**    If $F$ belongs to $\mathcal{G}$ and $g$ belongs to $\mathcal{L}$, then

$$\lim_{n \to \infty} \mu((F \circ T^n)\,g) = A_\mu(F)\,\mu(g).$$

**(M5)**    If $F$ belongs to $\mathcal{G}$, then

$$\lim_{n \to \infty} \sup_g \frac{1}{\mu(|g|)}\left|\mu((F \circ T^n)\,g) - A_\mu(F)\,\mu(g)\right| = 0,$$

where the supremum is taken over all the local observables $g$ in $\mathcal{L}$, except $g=0$.

The definition that we have discussed in the introduction with the number (5) is now relabeled **(M4)**. **(M3)** is a weaker version—much weaker, in fact, as discussed in Section 4.1. (Notice that in finite measure one can always assume that an observable has zero average; if not, it suffices to subtract its average, which is an operation of little consequence. The same cannot be done to local observables for an infinite-measure system.)

Finally, **(M5)** is a uniform version of **(M4)**, whose worth will be better explained in Proposition 3 below. For the moment, let us notice that the normalization by $\mu(|g|)$ (namely the $L^1$ norm of $g$) is rather natural, because the entire limit is linear in $g$.

## 2.6. Hierarchy of definitions

There are simple implications among the above notions:

**Proposition 2.**   Given the above definitions of global-global mixing, **(M2)** implies that the limit in **(M1)** holds true for all pairs $F,G$ in $\mathcal{G}$ such that $A_\mu(FoT^n)$ exists for all $n$ large enough (in particular, if the latter condition is verified for all pairs of global observables in $\mathcal{G}$, then **(M2)** $\Rightarrow$ **(M1)**).

As for the definitions of global-local mixing, **(M5)** $\Rightarrow$ **(M4)** $\Rightarrow$ **(M3)**.

*Proof.* The first statement is a direct consequence of the definition of the joint limit $\mu(V) \to \infty$, $n \to \infty$ in **(M2)**. The second chain of implication is obvious.                    Q.E.D.

Another, less obvious implication will be given in Section 4.1.

## 3. Examples

We now present a couple of examples of infinite-measure-preserving dynamical systems to which the scheme described in Section 2 can be naturally applied.

### 3.1. The random walk

A classical example of a dynamical system that is "uniformly chaotic" throughout its infinitely large phase space is the random walk. Not surprisingly, this was the main example in some of the previous literature on infinite mixing too, cf. [6]. There, a random walk was regarded as a Markov chain with an infinite state space. Here we give a more geometric representation.

Let us indicate with $a = (a_1, a_2, \ldots, a_d)$ a vector with $d$ integer components, equivalently, a site of the lattice $\mathbf{Z}^d$. Analogously for $b$. Let $\{p_a\}$ the collection of the transition probabilities defining a random walk; this means that the probability to jump from the site $a$ to the site $b$ is $p_{b-a}$. Let $\{a^{(j)}\}$ be an enumeration of the elements $a$ such that $p_a > 0$ (this enumeration can be finite or countably infinite), so $D = \{a^{(j)}\}$ is the set of the admissible directions for the random walk.

We partition the unit interval $I = [0,1)$, according to $D$, in the following way. Set $q_0 = 0$ and, recursively for $j > 0$, $q_j = q_{j-1} + p_{a(j)}$. Then set $I_j = [q_{j-1}, q_j)$. Clearly, $\{I_j\}$ is a partition of $I$. Notice that the Lebesgue measure of $I_j$, namely $p_{a(j)}$, is precisely the probability that the random walker makes the jump $a^{(j)}$.

Now for the definition of the dynamical system. Denote $M = \mathbf{Z}^d \times I$, which we may think of as the disjoint union of infinitely many copies of $I$. Let $\mu$ be the measure that coincides with the Lebesgue measure on each copy $\{a\} \times I$. In order to define $T: M \to M$, denote with $x = (a,y)$ an element of $M$; for $y$ in $[0,1)$, let $k = k(y)$ the unique positive integer such that $q_{k-1} \le y < q_k$ (equivalently, such that $y$ belongs to $I_k$). Then $T(x)$ is defined by

$$T(a,y) = \left( a + a^{(k)}, \frac{y - q_{k-1}}{q_k - q_{k-1}} \right).$$

Therefore $T$ acts on each $\{a\} \times I$ as a piecewise linear expanding map, and the branch of the map supported in $\{a\} \times I_k$ maps onto the interval $\{a + a^{(k)}\} \times I$. Fig. 1 shows this graphically. (The *cogniscenti* will recognize that $T$ is a Markov map relative to the partition $\{\{a\} \times I_k\}_{a,k}$.) It easy to verify that, choosing an initial condition $x$ at random from $\{b\} \times I$, with respect to $\mu$ (notice that $\mu(\{b\} \times I) = 1$), the sequence of sites visited by the trajectory $T^n(x)$ is a realization of the random walk introduced above.

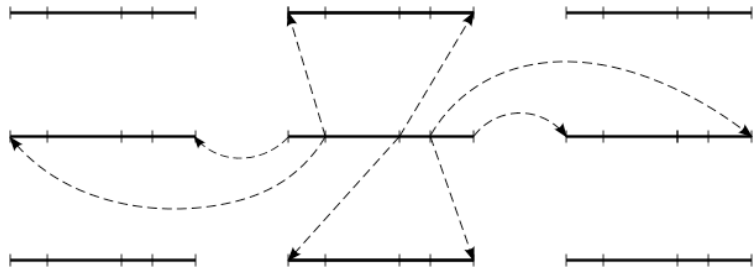

Fig. 1. A representation of the random walk map $T$. For every site of $\mathbf{Z}^d$ (in figure, $d=2$) there is a copy of the unit interval $I$, which is partitioned in a countable number of subintervals. $T$ maps each subinterval affinely onto one of the copies of $I$.

### *3.2. A specific one-dimensional case*

When *d=1*, the setting above can be further simplified, by means of the natural identification between *M* and ***R*** obtained by attaching the copies of *I* one after the other according to their order in ***Z***. To illustrate one specific example, consider the random walk in ***Z*** for which jumping to the left, remaining in the same position and jumping to the right have all probability *1/3*. Let *φ:**R**→**R*** be defined by *φ(x)=3x−1*, for *0≤ x< 1*, and *φ(x)=0*, otherwise. Then set $T(x)=\sum_a \varphi(x-a)+a$, where the sum is extended to all *a* in ***Z***. The graph of the resulting function of the real line onto itself is shown in Fig. 2. It is evident that *T* preserves *μ*, the Lebesgue measure on ***R***.

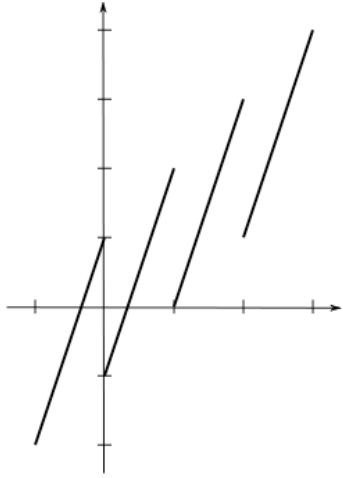

Fig. 2. The graph of the map *T* representing the random walk of Section 3.2.

### *3.3. Choosing the exhaustive family*

Coming back to the more general representation of a *d*-dimensional random walk, in order to verify the definitions **(M1)-(M5)** we need to choose the exhaustive family *V*, and then the classes of observables *G* and *L*. Let us start with *V*, leaving *G* and *L* to Section 4.2.

Let $B_{b,r}$ be the square in ***Z****$^d$* of center *b=(b₁,b₂,…,b_d)* and radius *r* in ***Z****$^+$*. In formula,

$$B_{b,r} = \left\{ a = (a_1, a_2, \ldots, a_d) \,\middle|\, b_i - r \le a_i \le b_i + r, \forall i = 1, 2, \ldots, d \right\}.$$

Set $V_{b,r}=B_{b,r}\times I \subset M$. We will refer to such sets as *boxes*.

In statistical mechanics, a typical way to define the infinite-volume limit—say, of a lattice system—is to select a box of radius *r* somewhere, for example with its center in the origin of the system, and then take the limit *r→∞*. If we were to follow this analogy strictly, we could define *V* as the collection of all the boxes $V_{0,r}$, with *r* in ***Z****$^+$* (in this case the limit *V↑M* would actually reduce to *r→∞*). We argue that this is not the best choice.

To illustrate the point, we use the specific example of the one-dimensional random walk given earlier, whose representation *(**R**,μ,T)* as a dynamical system we, by all means, want to classify as mixing. Suppose one considers the global observable *F:**R**→**R***, defined by *F(x)=−1*, for *x<0*, and *F(x)=1*, for *x≥0*. (Granted that we have not defined *G* yet, but this is actually part of the point, see later.) Clearly, *F* is bounded, and A_μ(F) exists and is equal to *0*. Also, by since the random walk has nearest-neighbor jumps, it is easy to see (see also Fig. 2) that *FoTⁿ(x)=−1*, for *x<−n*, and *FoTⁿ(x)=1*, for *x≥n*. In other words, a point *x* with *|x|>n* needs a time at least *n* to measure a different value of the observable *F*. Since the value of any global observable on *[−n,n]* is

irrelevant for the infinite-volume average, we have that $A_\mu((F \circ T^n)F) = I \neq (A_\mu(F))^2$. So, if $F$ belongs to $\mathcal{G}$, this excludes any possibility of global-global mixing, whether it be **(M1)** or **(M2)**.

This is an instance of a phenomenon that is in fact known in statistical mechanics, namely, the infinite-volume limit does not see *interface effects*. In our case, $F$ fails to be constant only around the set *{x=0}*, which acts as a lower-dimensional *interface* between the two values (or *phases*) of the observable *F*. Therefore, at least insofar as global-global mixing is concerned, *F* behaves like a constant observable, which is irrelevant in **(M1)-(M2)**.

So, what to do in this case? One possibility is to make sure that no observable giving rise to interface effects belongs to $\mathcal{G}$. But this might be quite cumbersome. A sounder and more elegant way to solve the problem is to realize that, since our system is translation invariant (with respect to the lattice $\mathbf{Z}^d$), then $\mathcal{V}$ should be too. Hence, a more reasonable choice for $\mathcal{V}$ is to comprise *all* the boxes $V_{b,r}$, with *b* in $\mathbf{Z}^d$ and *r* in $\mathbf{Z}^+$. In this way (because the infinite-volume limit is by definition uniform with respect to the location of *V*), observables like *F* are automatically excluded.

This demonstrates the point (that was also mentioned earlier) that the choice of $\mathcal{V}$ is very important and should reflect the features of a dynamical system. For example, if a system has symmetries or "quasi-symmetries" (i.e., transformations that change the system into a similar one), then $\mathcal{V}$ should be invariant for them. More considerations about this (and the all-important choice of $\mathcal{G}$) can be found in [3].

The proof that the $\mathcal{V}$ defined above verifies **(A1)** for our system is the same as the one found in the appendix of [3] for a technically different, but completely analogous, dynamical system.

### 3.4. The Farey map

Another class of infinite-measure-preserving dynamical systems on which it would be very interesting to test our definitions is given by the intermittent maps of the interval, i.e., maps *T:[0,1]→[0,1]* that have a neutral fixed point in *0*. To fix the ideas, we consider just one example, the celebrated Farey map (a.k.a., Parry-Daniels map), which is defined as follows: *T(0)=0* and, for *x* in *(0,1]*,

$$T(x) = \begin{cases} \dfrac{x}{1-x}, & x \in (0,1/2]; \\ \dfrac{1-x}{x}, & x \in (1/2,1]. \end{cases}$$

See the graph of *T* in Fig. 3. It is known that this map preserves the measure $\mu$ on *[0,1]* given by *dμ=dx/x* [12]. Therefore an infinite portion of the mass is concentrated around *0*, which plays the role of the point at infinity.

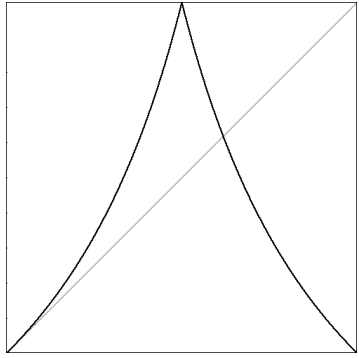

Fig. 3. The graph of the Farey map in the unit square. The gray line is the bisectrix of the first quadrant. The fact that the graph of the map is tangent to this line at the origin means that the dynamical system has a neutral fixed point there.

In this case a suitable choice for $\mathcal{V}$ is $\mathcal{V} = \{ [c,1] \mid 0 < c < 1 \}$, because there are no evident symmetries in the system and, in a sense, there is only one way to go to infinity. (In truth, arguments could be made in favor of the choice $\mathcal{V}' = \{ [c,d] \mid 0 < c < d \leq 1 \}$, which also has the advantage of restricting *a priori* the class of global observables by making it harder to verify **(A3)**. We do not delve into this issue here.)

Let us verify **(A1)** for $\mathcal{V}$. Set $V=[c,1]$ and $x_c=c/(1+c)$. It is easy to check that $T^{-1}V=[x_c , 1-x_c ]$, whence

$$\mu(T^{-1}V \, \Delta \, V) = \mu([x_c , c] \cup [1-x_c , 1]) = \log \frac{c}{x_c} + \log \frac{1}{1-x_c} = 2\log(1+c),$$

which tends to *0*, as $c \to 0^+$, equivalently, as $V \uparrow [0,1]$.

## 4. General and specific results

In this section we list some results that can be proven about the definitions **(M1)-(M5)**. We start with general propositions and then specialize to the examples of Section 3.

### 4.1. General results

Proposition 2 ranks the two definitions of global-global mixing and the three definitions of global-local mixing in order of strength. There is an additional relation between the two different concepts: **(M5)**, the strongest form of global-local mixing implies **(M2)**, the "strongest" form of global-global mixing, at least for a very general class of global observables:

**Proposition 3.** Suppose that every $F$ in $\mathcal{G}$ can be written as $F=\sum_j f_j$ (the sum is on all $j$ in $N$), with $f_j$ in $\mathcal{L}$, and, for every $V$ in $\mathcal{V}$, there exists a finite set $J_V \subset N$ such that,

$$\left\| F \mathbf{1}_V - \sum_{j \in J_V} f_j \right\|_1 = o(\mu(V)); \tag{7}$$

$$\sum_{j \in J_V} \left\| f_j \right\|_1 = O(\mu(V)) \tag{8}$$

(where $\mathbf{1}_V$ is the indicator function of $V$ and $\| \cdot \|_1 = \mu(|\cdot|)$ is the $L^1$ norm for functions of $(M, \Sigma, \mu)$). Then **(M5) ⇒ (M2)**.

The proof of Proposition 3 can be found in [3]. Conditions (7)-(8) are less complicated than they appear. One should think of the very common situation in which $M$ admits a partition of the unity, $\sum_j \psi_j = 1$, where $\psi_j$ are nonnegative integrable functions which are roughly translations of each other. In many such cases, $f_j = F \psi_j$ will verify both of the above conditions.

To state the next two results, we need a couple of standard definitions from ergodic theory [1].

**Definition 4.** A measure-preserving dynamical system $(M, \Sigma, \mu, T)$ is called *exact* if $\cap_{n>0} T^{-n} \Sigma = \Omega$, where $\Omega$ is the so-called *trivial σ-algebra*, i.e., the σ-algebra that only contains the $\mu$-null sets and their complements.

**Definition 5.** An invertible measure-preserving dynamical system $(M, \Sigma, \mu, T)$ is called *K-mixing* if there exists a σ-algebra $\Gamma \subset \Sigma$ such that:
- $\Gamma \subset T\Gamma$;
- $\sigma(\cup_{n>0} T^n \Gamma) = \Sigma$ mod $\mu$, where the left-hand side is the σ-algebra generated by all the σ-algebras $T^n \Gamma$, $n>0$;
- $\cap_{n>0} T^{-n} \Sigma = \Omega$, where $\Omega$ is the trivial σ-algebra, as defined above.

Morally speaking, Definitions 4 and 5 address the same property of a dynamical system (exactness being applied to non-invertible maps and K-mixing to invertible ones), a property that in finite ergodic theory implies *mixing at all orders*. We do not delve on this point here.

**Proposition 6.** If $(M, \Sigma, \mu, T)$ is exact then **(M3)** holds with any choice of $\mathcal{G}$ and $\mathcal{L}$ (that satisfy **(A1)-(A4)**).

*Proof.* Define the Perron-Frobenius operator $P$ via the identity $\mu((F \circ T)g) = \mu(F(Pg))$, where $F$ ranges in $L^\infty(M, \Sigma, \mu)$ and $g$ in $L^1(M, \Sigma, \mu)$. This is morally the same as the operator $Q$ introduced in Section 1, except that it acts on $L^1$ functions. A famous theorem by Lin [13] states that, if $T$ is exact and $\mu(g) = 0$, then $\|P^n g\|_1 \to 0$, as $n \to \infty$. Therefore $|\mu((F \circ T^n)g)| \leq \|F\|_\infty \|P^n g\|_1 \to 0$, which is **(M3)**.                                                                                                        Q.E.D.

**Proposition 7.** Let $(M, \Sigma, \mu, T)$ be K-mixing. For $m>0$, define $\mathcal{G}_m = L^\infty(M, T^m \Sigma, \mu)$ (which is the space of the $\mu$-essentially bounded functions that are measurable with respect to $T^m \Sigma$) and assume that $\mathcal{G}$ is contained in the closure of $\cup_{m>0} \mathcal{G}_m$, relative to the norm $\| \cdot \|_\infty$. Then **(M3)** holds with any choice of $\mathcal{L}$.

The proof of Proposition 7 will be given elsewhere.

The above propositions are important both for their great generality and because they clarify how **(M3)** is a much weaker result than **(M4)**. The former can hold without the case-by-case assumptions one must make on $\mathcal{G}$ to make sure that—using the notation of Section 1—$Q^n \mu_g(F)$ is a converging sequence, for all $F$ in $\mathcal{G}$ (in which case **(M4)** is equivalent to the limit being $A_\mu(F)$). **(M3)** only requires that, if $f$ and $g$ are both probability densities relative to $\mu$, then $Q^n \mu_f(F) - Q^n \mu_g(F) \to 0$.

### 4.2. The random walk

For the dynamical system of Section 3.1, which represent a random walk in $\mathbf{Z}^d$, we define a countable number of classes of global and local observables, as we can prove stronger or weaker results depending on the classes $\mathcal{G}$, $\mathcal{L}$ that are being considered.

Let $\Pi = \{\{a\} \times I\}_a$, that is, $\Pi$ is the partition of $M$ given by the infinitely many copies of $I$; cf. Section 3.1. For $m>0$, denote by $\Gamma_m$ the $\sigma$-algebra generated by the partitions $\Pi$, $T^{-1}\Pi$, …, $T^{-m}\Pi$, and define $\mathcal{G}_m = \{F$ in $L^\infty(M, \Gamma_m, \mu) \mid A_\mu(F)$ exists$\}$. To get an idea of the functions we are dealing with here, let us consider the one-dimensional example of Section 3.2: a function that is measurable with respect to $\Gamma_m$ is constant on the intervals $[k3^{-m}, (k+1)3^{-m})$, with $k$ in $\mathbf{Z}$. In the general case, if the random walk is non-trivial (i.e., $D$ contains more than one element), $\Gamma_m$ is a finer and finer $\sigma$-algebra of $M$, as $m \to \infty$, and the elements of its generating partition have diameters that vanish exponentially. Define $\mathcal{G}$ as the closure of $\cup_{m>0} \mathcal{G}_m$ in the $L^\infty$ norm. This is a pretty large class of global observables, which includes, for example, all uniformly continuous bounded functions for which the infinite-volume average exists (this is easily implied by the considerations above).

Finally, set $\mathcal{L}_m = L^1(M, \Gamma_m, \mu)$ and $\mathcal{L} = L^1(M, \Sigma, \mu)$, where $\Sigma$ is the natural $\sigma$-algebra for $M$, i.e., the one generated by the standard $\sigma$-algebras defined on the many copies of $I=[0,1)$. Clearly, $\mathcal{L}$ is the closure of $\cup_{m>0} \mathcal{L}_m$ in the $L^1$ norm.

To state our theorem we need a standard definition for random walks [14].

**Definition 8.** Given the set $D = \{a^{(j)}\}$ of the admissible jumps for the random walk (cf. Section 3.1), set $\mathbf{Y} = \text{span}_{\mathbf{Z}}\{a^{(j)} - a^{(1)}\}_{j>1}$, which is the subgroup of $\mathbf{Z}^d$ made up of the finite linear combinations of the vectors $a^{(j)} - a^{(1)}$, for $j>1$, with coefficients in $\mathbf{Z}$. (It can be seen [3] that $\mathbf{Y}$ does not depend on the ordering put on $D$). The random walk is called *strongly aperiodic* if $\mathbf{Y} = \mathbf{Z}^d$.

**Theorem 9.** Let $(M, \mu, T)$ be the dynamical system defined in Section 3.1. Set $\nu = \max\{2, [d/2]+1\}$, where $[\cdot]$ is the integer part of a positive number, and suppose that the probability distribution of the jumps has a finite $\nu^{\text{th}}$ momentum, i.e.,

$$\sum_{a \in \mathbf{Z}^d} |a|^\nu p_a < \infty.$$

If the associated random walk is strongly aperiodic, as per Definition 8, the system is mixing in the following senses:

- **(M5)** relative to $\mathcal{G}_m$ and $\mathcal{L}_m$, for all $m$ in $N$;
- **(M4)-(M3)** relative to $\mathcal{G}$ and $\mathcal{L}$;
- **(M2)** relative to $\mathcal{G}$;
- **(M1)** relative to $\mathcal{G}_m$, for all $m$ in $N$, provided that $F$ is $\mathbf{Z}^d$-periodic, i.e. $F(a,y) = F(0,y)$, for all $a$ in $\mathbf{Z}^d$ and y in $I$.

The proof of this theorem is a corollary of the corresponding theorem in [3], stated for the dynamical system that is used there.

We remark that the hypothesis of strong aperiodicity is essential, and one should not expect to have mixing without it. To see this, consider for example a case where $a^{(1)}=0$ and $Y \neq Z^d$. Then one can consider the global observable $F$ such that $F(a,y)=1$, for all $a$ in $Y$, and $F(a,y)=0$, otherwise. Clearly, **(A2)-(A3)** are verified, but no mixing is possible since the two level sets of $F$ are invariant for the map $T$.

### 4.3. The Farey map

For the Farey map, defined in Section 3.4, we make the most general (which in this case means the smallest) choice for the exhaustive family, namely, $\mathcal{V} = \{[c,1] \mid 0 < c < 1\}$. A reasonable choice for $\mathcal{G}$ could then be the continuous functions $F:[0,1] \to R$. Obviously **(A2)-(A3)** are verified. Specifically, we have the following obvious result.

**Lemma 10.** If $F$ belongs to $\mathcal{G}$, then $A_\mu(F)=F(0)$. If $G$ also belongs to $\mathcal{G}$, then $FG$ belongs to $\mathcal{G}$ and $A_\mu(FG)= A_\mu(F)A_\mu(G)$.

The second statement above corresponds to the ring property mentioned in Section 2.4, and its proof is a trivial consequence of the first statement, which is in turn obvious, considering that the overwhelming majority (in fact, the infinite part) of the measure $\mu$ is concentrated around $0$.

As for the local observables, let us make the most general choice, namely, $\mathcal{L} = L^1([0,1],\mu)$.

**Theorem 11.** Let $([0,1],\mu,T)$ be the Farey map defined in Section 3.4, and $\mathcal{G}, \mathcal{L}$ as described above. Then **(M1)-(M4)** hold true.

*Proof.* Since, for any fixed $n$ in $N$, $T^n$ is continuous in a neighborhood of $0$ and $T^n(0)=0$, it is clear that $A_\mu((F \circ T^n)G)=F(0)G(0)=A_\mu(F)A_\mu(G)$, for all $F,G$ in $\mathcal{G}$. This immediately implies **(M1)** and **(M2)**.

As for the global-local mixing, let us introduce other classes of global observables. For $0<a<1$, define $\mathcal{G}_a = \{F$ in $\mathcal{G} \mid F$ is constant on $[0,a]\}$. It is a consequence of a result by Isola on the scaling rate for the Farey map [15] that, for all bounded functions $f,g$ that are supported in $[0,a]$, $\mu((f \circ T^n)g) \to 0$, as $n \to \infty$. Since any $F$ in $\mathcal{G}_a$ can be rewritten as $F(x)=F(0)+f(x)$, with $f$ bounded and supported in $[0,a]$, the above implies that, for such $F$ and any $g$ bounded and supported in $[0,a]$, $\mu((F \circ T^n)g) \to A_\mu(F)\mu(g)$.

By density arguments (using the $L^1$ norm for local observables and the $L^\infty$ norm for global observables) this implies **(M4)** for local observables in $\mathcal{L}$ and global observables in the $L^\infty$-closure of $\cup_{a>0} \mathcal{G}_a$, which is $\mathcal{G}$. **(M3)** follows.        Q.E.D.

Theorem 11, together with an analog of Lemma 10, can as easily be proved for the class $\mathcal{G}_0$ of all functions $F:[0,1] \to R$ which have a limit at $0$. Denoting such limit by $F(0^+)$, it is clear that $A_\mu(F)=F(0^+)$. All is based on the fact that $\mu$ is infinitely concentrated around $0$, so, as far as the infinite-volume average is concerned, any such global observable is the same as a constant function, and "decorrelation" for constant functions is trivial.

Therefore $\mathcal{G}$, as well as $\mathcal{G}_0$, is not a very relevant class of global observables. One should consider functions that are non-constant around the "point at infinity". An extreme choice might be $\mathcal{G}'$, the largest class verifying **(A2)-(A3)** (in other words, $\mathcal{G}'$ comprises all essentially bounded functions for which the infinite-volume average exists). Then the results change.

**Theorem 12.** For the Farey map $([0,1],\mu,T)$ endowed with the classes $\mathcal{G}'$ and $\mathcal{L}$ for the global and local observables, respectively, **(M1)-(M2)** are false and **(M3)** is true.

*Sketch of the proof.* **(M3)** follows from Proposition 6 and the exactness of the map [12]. The falseness of **(M1)-(M2)** is a consequence of the fact that, although any neighborhood of *0* carries the infinite majority of the measure, very little "chaos" happens there. If one takes a global observable whose variation around *0* has a certain regularity, e.g., $F(x) = \sin(e^{-x})$, it can be seen that, for all *n* in *N*, $FoT^n(x) \approx F(x)$, in a sense that we do not fully specify here but that implies that $A_\mu((FoT^n)F) = A_\mu(F^2) > 0$. On the other hand, $A_\mu(F) = 0$.                                                          Q.E.D.

A more detailed proof of Theorem 12 will be presented elsewhere. We conclude by noting that, in this case, **(M4)** is a rather hard problem, which is yet to proved or disproved.

### Acknowledgements

I would like to thank Stefano Isola for many instructive discussions on the subject.